\documentstyle[12pt]{article}
\input epsf
\def \b{{\cal B}}
\def \bk{\overline{K}^0}
\def \G{{\rm GeV}}
\def \ite{{\it et al.}}
\def \k{K^0}
\def \cn{Collaboration}
\def \beq{\begin{equation}}
\def \eeq{\end{equation}}

\begin{document}
\rightline{EFI-97-58}
\rightline{hep-ph/9801201}
\rightline{December 1997}
\bigskip
\bigskip
\centerline{{\bf $B$ PHYSICS -- A THEORETICAL REVIEW}
\footnote{Presented at Beauty '97 -- Fifth International Workshop on
$B$-Physics at Hadron Machines, UCLA, October 13--17, 1997. To be published in
Nuclear Instruments and Methods.}} 
\bigskip
\centerline{\it Jonathan L. Rosner}
\centerline{\it Enrico Fermi Institute and Department of Physics}
\centerline{\it University of Chicago, Chicago, IL 60637}
\bigskip
\centerline{\bf ABSTRACT}
\medskip
\begin{quote}
This overview of what we can hope to learn from high-statistics experiments in
$B$ physics in the next few years includes: (a) a review of parameters of the
Cabibbo-Kobayashi-Maskawa (CKM) Matrix; (b) direct determination of magnitudes
of CKM elements; (c) forthcoming information from studies of kaons; (d) CP
violation in $B$ decays; (e) aspects of rate measurements; (f) the role of
charm-anticharm annihilation; (g) remarks on tagging; and (h) effects beyond
the standard model. 
\end{quote}
\bigskip

\centerline{\bf I.  INTRODUCTION}
\bigskip

The violation of CP symmetry in the decays of neutral kaons \cite{CCFT} remains
a mystery more than thirty years after its discovery.  A candidate theory for
this violation, involving phases in the Cabibbo-Kobayashi-Maskawa (CKM) matrix
\cite{Cab,KM},  predicts large CP-violating effects in the decays of $B$
mesons.  As a result, a massive experimental assault is under way to study
those decays.  In the present review we discuss some types of studies that will
be possible in high-statistics experiments, whether by advances in background
reduction in hadronic experiments or by increased luminosity in
electron-positron collisions.  This report is in part a summary of other
theoretical contributions at the {\it Beauty '97}~Workshop, which should be
consulted for details. 

We begin in Section II with a brief overview of the CKM parameters and why we
care about their precise values.  Section III is devoted to direct measurements
of magnitudes of CKM elements. Experiments with kaons, some of which are very
close to presenting new results, will provide partial information (Sec.~IV). 
Various ways of detecting CP violation in $B$ decays exist (Sec.~V); we
concentrate on measurements of rates for rare processes (Sec.~VI) which do not
require time-dependent studies.  Some brief remarks are made in Section VII
about the role of charm-anticharm annihilation in $B$ decays, and in Section
VIII about progress in identifying the flavor of a neutral $B$ meson at time of
production.  Some tests of physics beyond the standard model are mentioned in
Section IX, while Section X concludes. 
\bigskip

\centerline{\bf II.  WHY WE CARE ABOUT CKM PARAMETERS}
\bigskip

We give an abbreviated version of a description which may be found in more
detail elsewhere \cite{JRKEK}. The CKM matrix element $V_{ij}$ describes the
charge-changing transition of a left-handed down-type quark $j$ to a
left-handed up-type quark $i$.  A parametrization sufficiently accurate for
present purposes is that of Wolfenstein \cite{WP}: 

\beq
V \approx \left[ \matrix{1 - \lambda^2/2 & \lambda & A \lambda^3 (\rho -
i \eta) \cr
- \lambda & 1 - \lambda^2 /2 & A \lambda^2 \cr
A \lambda^3 (1 - \rho - i \eta) & - A \lambda^2 & 1 \cr } \right].
\eeq
The quantity $\lambda = 0.2205 \pm 0.0018 = \sin \theta_c$ \cite{strange,PDG}
expresses the suppression of $s \to u$ decays with respect to $d \to u$ decays
\cite{Cab,GL}. This parameter describes the $u,~d,~s$, and $c$ couplings via
the upper left $2 \times 2$ submatrix of $V$ \cite{charm}. 

When a third family of quarks is added, three more parameters are needed.  One
may express them in terms of (1) the strength characterizing $b \to c$ decays,
$A \lambda^2 = 0.0393 \pm 0.0028$ \cite{Gib,AL} so that $A = 0.81 \pm 0.06$
(see \cite{MN}, \cite{ALK} for slightly different values); (2) the magnitude of
the $b \to u$ transition element measured in charmless $b$ decays,
$V_{ub}/V_{cb} = 0.08 \pm 0.02$ \cite{Gib} so that 
\beq \label{eqn:Vubcon}
(\rho^2 + \eta^2)^{1/2} = 0.36 \pm 0.09~~;
\eeq
(3) the phase of $V_{ub} = A \lambda^3 (\rho - i \eta)$. Unitarity of the
CKM matrix implies that the scalar product of the complex conjugate of a row
with any other row should vanish, e.g., 
\beq \label{eqn:ur}
V_{ud}^* V_{td} + V_{us}^* V_{ts} + V_{ub}^* V_{tb} = 0 ~~~.
\eeq
Since $V_{ud}^* \approx 1,~V_{us}^* \approx \lambda,~V_{ts} \approx - A
\lambda^2$, and $V_{tb} \approx 1$ we have $V_{td} + V_{ub}^* = A \lambda^3$.
Dividing (\ref{eqn:ur}) by $A \lambda^3$, since $V_{ub}^*/ A \lambda^3 = \rho +
i \eta, ~~ V_{td} / A \lambda^3 = 1 - \rho - i \eta$, one obtains the triangle
shown in Fig.~1. Here the angles $\alpha, \beta$, and $\gamma$ are defined as
in \cite{NQ}. The value of $V_{ub}^* / A \lambda^3$ may then be depicted as a
point in the $(\rho,\eta)$ plane. To resolve the major remaining ambiguity in
the determination of the CKM matrix elements, i.e., the phase of $V_{ub}$, the
shape of the unitarity triangle, or the value of $V_{td}$, one must resort to
indirect means, which involve loop diagrams. 
 
\begin{figure}
\centerline{\epsfysize = 1.6in \epsffile {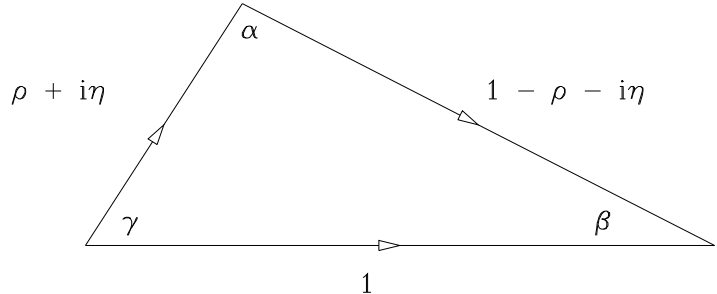}}
\caption{Unitarity triangle for CKM elements. We show in the complex plane the
relation (\protect\ref{eqn:ur}) divided by the normalizing factor $A
\lambda^3$.} 
\end{figure}

Box diagrams involving $u,~c,$ and $t$ in loops contribute to the virtual $b
\bar d \leftrightarrow d \bar b$ transitions which mix $\bar B^0$ and $B^0$.
The leading contribution at high internal momentum in these diagrams cancels as
a consequence of (\ref{eqn:ur}).  The remaining contribution is dominated by
the top quark since all products of CKM elements $V_{qb} V_{qd}^*$ are of order
$\lambda^3$ while $m_t \gg m_c,~m_u$.

In the calculation of the effect of the box diagram \cite{IL} one needs several
parameters, the most recent values of which are $m_t = 175.5 \pm 5.5~\G/c^2$
\cite{twm}, $M_W = 80.40 \pm 0.08~\G/c^2$ \cite{twm}, $m_B = 5.279~\G/c^2$ (see
\cite{PDG}), and $f_B \sqrt{B_B} = 200 \pm 40$ MeV \cite{AL,MN}.  Here $f_B$ is
the $B$ meson decay constant, defined so that the matrix element of the weak
axial-vector current $A_\mu \equiv \bar b \gamma_\mu \gamma_5 d$ between a
$B^0$ meson and the vacuum is $\langle 0 | A_\mu | B^0(p) \rangle = i p_\mu
f_B$.  The factor $B_B$ expresses the degree to which the box diagrams provide
the contribution to $B - \bar B$ mixing.  The result involves a QCD correction
$\eta_B = 0.55$ \cite{QCDB}.  The mixing amplitude \cite{Bmix} is $\Delta m_d
= 0.472 \pm 0.018~{\rm ps}^{-1}$, where the subscript refers to the mixing
between $B^0 \equiv \bar b d$ and $\bar B^0 \equiv b \bar d$.  The
corresponding estimate of $|V_{td}|$ (see Ref.~\cite{JRKEK} for more details)
leads, once we factor out a term $A \lambda^3$, to the constraint \cite{GKR} 
\beq \label{eqn:Bmixcon}
|1 - \rho - i \eta | = 1.01 \pm 0.22~~~.
\eeq  

A key constraint comes from CP-violating $\k$--$\bk$ mixing.  One can
parametrize the mass eigenstates as $K_S \simeq K_1 + \epsilon K_2,~K_L \simeq
K_2 + \epsilon K_1$, where $|\epsilon| \simeq 2 \times 10^{-3}$ and the phase
of $\epsilon$ turns out to be about $\pi/4$.  The parameter $\epsilon$ encodes
all current knowledge about CP violation in the neutral kaon system.  Its
origin is still the subject of hypothesis. One possibility, proposed \cite{sw}
immediately after the discovery and still not excluded, is a ``superweak''
CP-violating interaction which directly mixes $\k = d \bar s$ and $\bk = s \bar
d$. The presence of three quark families \cite{KM} poses another opportunity
for explaining CP violation through box diagrams involving $u,~c$, and $t$
quarks. With three quark families, phases in complex coupling coefficients
cannot be removed by redefinition of quark phases.  Within some approximations
\cite{JRCP}, the parameter $\epsilon$ is directly proportional to the imaginary
part of the mixing amplitude.  Its magnitude was given for arbitrary $m_t$ in
Ref.~\cite{IL} and implies the constraint \cite{JRKEK} 
\beq \label{eqn:Kmixcon}
\eta(1 - \rho + 0.44) = 0.51 \pm 0.18~~~,
\eeq
where the term $1 - \rho$ corresponds to the loop diagram with two top quarks,
and the term 0.44 corresponds to the additional contribution of charmed quarks.
The major source of error on the right-hand side is the uncertainty in the
parameter $A \equiv V_{cb}/\lambda^2$. Eq.~(\ref{eqn:Kmixcon}) can be plotted
in the $(\rho,\eta)$ plane as a band bounded by hyperbolae with foci at
(1.44,0).

The most recent constraint is due to a lower bound on the mixing parameter
$\Delta m_s$ for $B_s^0$--$\bar B^0_s$ mixing: $\Delta m_s > 10.2$ ps$^{-1}$
\cite{Bmix}.  The loop diagrams for $\Delta m_s$ are dominated by the CKM
element $V_{ts}$, in contrast to those for $\Delta m_d$ which are dominated by
$V_{td}$.  Thus a measurement of $\Delta m_s/\Delta m_d$ implies a limit on
$|V_{ts}/V_{td}|$.  The result \cite{Bmix} is $|V_{ts}/V_{td}| > 3.8$,
implying $|1-\rho-i\eta| < 1.19$.  This excludes a small portion of the region
allowed by the $B^0$--$\bar B^0$ mixing constraint (\ref{eqn:Bmixcon}). 

The $(\rho,\eta)$ region allowed by all the above constraints is plotted in
Fig.~2.  The boundaries are dominated by theoretical uncertainties. 

\begin{figure}
\centerline{\epsfysize = 3.3in \epsffile {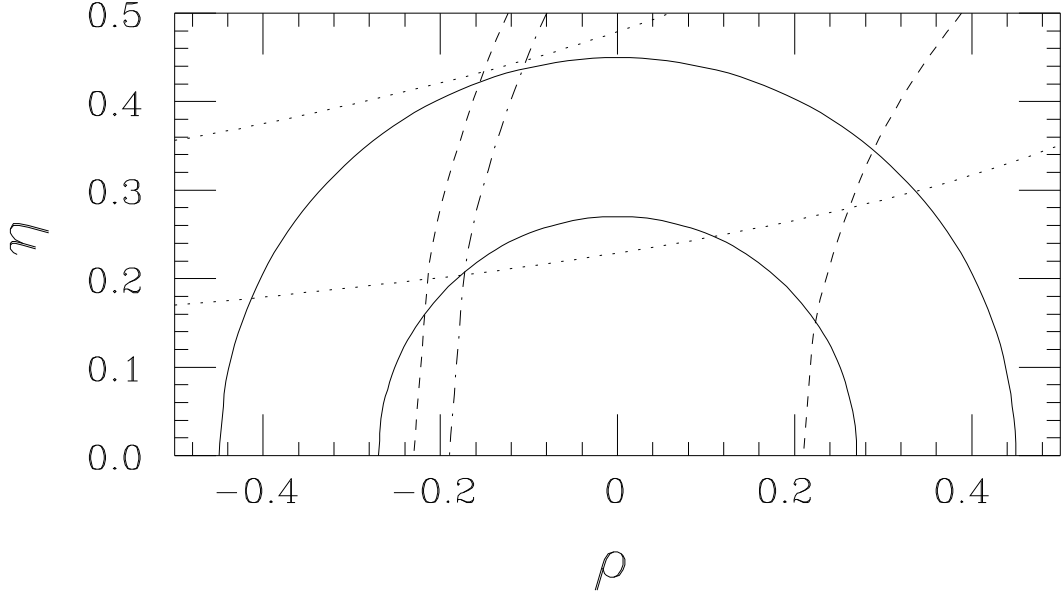}}
\caption{Region in the $(\rho,\eta)$ plane allowed by constraints on
$|V_{ub}/V_{cb}|$ (solid semicircles), $B^0$--$\bar B^0$ mixing (dashed
semicircles), CP-violating $K$--$\bar K$ mixing (dotted hyperbolae),
and $B_s^0$--$\bar B_s^0$ mixing (to the right of the dot-dashed semicircle).}
\end{figure}
 
A region centered about $\rho \simeq 0.05$, $\eta \simeq 0.35$ is permitted.
Nonetheless, the CP violation seen in kaons could be due to an entirely
different source, such as a superweak mixing of $K^0$ and $\bar K^0$ \cite{sw}.
In that case one could still accommodate $\eta = 0$, and hence a real CKM
matrix, by going slightly outside the bounds based on $|V_{ub}/V_{cb}|$ or
$B$--$\bar B$ mixing.  This circumstance is illustrated by a considerably more
permissive $(\rho,\eta)$ plot presented in Ref.~\cite{Nir}, for which the
$B$--$\bar B$ mixing constraint entails only $0.65 < |1-\rho-i\eta| < 1.63$.
Even with this relaxed constraint, the range of values of $\sin(2 \beta)$ is
quite restricted, lying roughly between 0.4 and 0.9 (assuming that $\sin(2
\beta) > 0$ as implied by the value of $\epsilon$!) Thus, there is much
potential for uncovering new physics simply by observing a value of $\sin(2
\beta)$ outside the range 0.4--0.9.  Such new physics is most likely to be
manifested most prominently (though not exclusively) through additional
contributions to mixing, which can affect the determination of $\sin(2 \beta)$
via a CP asymmetry in the $B^0 \to J/\psi K_S$ mode. 

The $(\rho,\eta)$ plot can be useful in several ways.  (1) It permits
anticipation of standard model expectations for $K \to \pi \nu \bar \nu$, $B$
decays, and other experiments.  (2) It permits one to exclude a superweak model
in which the CP-violating asymmetries in $B \to J/\psi K_S$ and $B \to \pi^+
\pi^-$ are equal and opposite \cite{BW}.  (3) It allows one to constrain
schemes in which CKM matrix elements are related to quark masses, typically by
means of a series of discrete assumptions. (4) It allows one to expose
inconsistencies in the standard CKM through overconstrained measurements of the
sides and angles of the unitarity triangle.  New-physics effects are seen
typically (though not exclusively) in mixing and penguin amplitudes (e.g.,
those dominating the $b \to s \gamma$ and $b \to s \ell^+ \ell^-$ subprocesses)
before showing up in direct decays, because loop effects are more sensitive to
these effects. 
\newpage

\centerline{\bf III.  MAGNITUDES AND RATIOS OF CKM MATRIX ELEMENTS}
\bigskip

\noindent
{\bf A.  $|V_{cb}|$}.
Improvement of accuracy in $|V_{cb}|$ is particularly important since the
dominant contribution to $\epsilon$ in CP-violating $K$--$\bar K$ mixing
behaves as $|V_{cb}|^4$.  In {\it inclusive} decays $B \to X_c \ell \nu_\ell$
\cite{Falk} it will be useful to measure $M(X_c)$ and moments $\langle
E_\ell^n \rangle$ of the lepton energy $E_\ell$.  The fact that $m_b - m_c$
is constrained to lie in a narrow range between 3.34 and 3.4 GeV$/c^2$ helps
to reduce uncertainty due to the the overall scale of $m_b$ and $m_c$
\cite{TASI}.  One hopes for an eventual determination of $|V_{cb}|$ to an
error of 5\%.

Among {\it exclusive} decays the process $B \to D^* \ell \nu_\ell$ holds the
best hope of yielding an accurate value of $|V_{cb}|$.  One uncertainty in the
corresponding form factors \cite{Wise} is associated with the contribution of
virtual pions: e.g., $B \to B^* \pi \to D \pi \ell \nu_\ell \to D^* \ell
\nu_\ell$.  The $D^*D \pi$ coupling constant $g$ enters in this calculation and
could in principle be measured:  $\Gamma(D^{*+} \to D^0 \pi^+) = (0.2 ~{\rm
MeV})g^2$.  One expects \cite{Dstar} $g \simeq 1/2$, posing a real challenge to
experimentalists.  One might be able to ``calibrate'' the $D^*$ widths if it
were possible to calculate the rate for $D^{*0} \to D^0 \gamma$ accurately. 
\bigskip

\noindent
{\bf B.  $|V_{ub}/V_{cb}|$}.
Information on $|V_{ub}/V_{cb}|$ constrains $(\rho^2 + \eta^2)^{1/2}$ and bears
improvement, as shown in Fig.~2.  {\it Inclusive} decays \cite{Falk} $B \to X
\ell \nu_\ell$ with $M_X < m_D$ would allow one to study a larger part of the
spectrum in charmless semileptonic decays than is possible with the current
method, which looks at the ``tail of the elephant'' by measuring the lepton
energy spectrum beyond the charm endpoint. 

{\it Exclusive} decays can improve information on $|V_{ub}/V_{cb}|$ \cite{Wise}
by studying $B \to \rho \ell \nu_\ell$ and relating its form factors to those
in the process $D \to \rho \ell \nu_\ell$, which in turn can be related by
flavor SU(3) to $D \to K^* \ell \nu_\ell$.  It may also be possible to gain
adequate understanding of the form factor in $B \to \pi \ell \nu_\ell$
\cite{Ball}, for which a measurement at the 25\% level now exists \cite{Semi}.
One foresees an eventual determination of $|V_{ub}/V_{cb}|$ to 10\%. 
\bigskip

\noindent
{\bf C.  $|V_{ts}/V_{td}|$}.
As mentioned, the $(\rho,\eta)$ limits based on $x_s \equiv (\Delta m/\Gamma)
_{B_s}$ are beginning to encroach upon the otherwise-allowed region.  The
limit $\Delta m_s/\Delta m_d > 21.2$ (95\% c.l.) \cite{Bmix} corresponds
to $\Delta m_s > 10.2$ ps$^{-1}$ and $x_s > 16$.  The ratio
\beq
\sqrt{ \frac{\Delta m_s}{\Delta m_d} } \simeq \frac{f_{B_s}}{f_B} \left|
\frac{V_{ts}}{V_{td}} \right|~~~,
\eeq
combined with the quark-model upper limit $f_{B_s}/f_B < 1.25$ \cite{QMFB}
(consistent with lattice gauge estimates), implies $|V_{ts}/V_{td}| > 3.8$ and
$|1 - \rho - i \eta| < 1.19$ as shown in Fig.~2.  The allowed region permits
$|1 - \rho - i \eta| > 0.6$, or $x_s < (1.19/0.6)^2 \simeq 64$.
\bigskip

\noindent
{\bf D. $|V_{ub}/V_{td}|$}.
A direct measurement of the combination $f_B |V_{ub}|$ is possible through
the leptonic decay of the $B^+$:
\beq
\Gamma \left( B^+ \to \left\{ \begin{array}{c} \mu^+ \nu_\mu \\ \tau^+ \nu_\tau
\end{array} \right\}\right)  = \left\{ \begin{array}{c} 1.1 \times 10^{-7} \\
2.5 \times 10^{-5} \end{array} \right\} \left( \frac{f_B}{f_\pi} \right)^2
\left| \frac{V_{ub}}{0.003} \right|^2~~~.
\eeq
Given enough events, the $\mu^+ \nu_\mu$ mode is probably easier to detect
because of its clean signature.  In the ratio $\Gamma(B^+ \to \mu^+ \nu_\mu)
/\Delta m_d$, the factor $f_B$ cancels, leaving us with the ratio $R^2 \equiv
|V_{ub}/V_{td}|^2$.  Contours of constant $R$ \cite{HR} are circles in the
$(\rho,\eta)$ plane with centers $(\rho_0,\eta_0) = (R^2/[R^2-1],0)$ and
radii $R/|1-R^2|$.  For $(\rho,\eta) = (0,0.36)$ a 10\% measurement of $R^2$
(achievable with 100 observed $B^+ \to \mu^+ \nu_\mu$ decays) gives $\eta$ to
$\pm 0.02$.  For $\b(B^+ \to \mu^+ \nu_\mu) = 2 \times 10^{-7}$ one then
needs 500 million $B^+$, or about 500 million $B \bar B$ pairs produced at the
$\Upsilon(4S)$.
\bigskip

\centerline{\bf IV.  KAON INFORMATION}
\bigskip

The parameter $\epsilon'/\epsilon$, a measure of direct CP violation, is
probed in the double ratio
\beq \label{eqn:Krat}
\frac{\Gamma(K_L \to \pi^0 \pi^0)}{\Gamma(K_S \to \pi^0 \pi^0)} /
\frac{\Gamma(K_L \to \pi^+ \pi^-)}{\Gamma(K_S \to \pi^+ \pi^-)}
= 1 - 6~{\rm Re} \frac{\epsilon'}{\epsilon}~~~.
\eeq
In principle $\epsilon'/\epsilon$ is proportional to $\eta$, though there are
significant corrections due, for example, to electroweak penguins, and hadronic
matrix elements are very uncertain.  A likely range is $\epsilon'/\epsilon
= (0 - 1) \times 10^{-3}$ \cite{eps}.  If $\epsilon'/\epsilon \ne 0$ the
superweak theory \cite{sw} will finally have been disproved.

The decay $K^+ \to \pi^+ \nu \bar \nu$ is sensitive to loop processes, mainly
to $V_{td}$ but also to a non-negligible charm contribution.  Thus, it roughly
measures the parameter $|1.4 - \rho - i \eta|$. The standard prediction
\cite{BB} is $\b(K^+ \to \pi^+ \nu \bar \nu) \simeq 10^{-10} \times 2^{\pm 1}$.
After a search of several years, Brookhaven Experiment E-787 has finally seen
one event for this process \cite{BNL}, corresponding to a branching ratio of
$(4.2^{+9.7}_{-3.5}) \times 10^{-10}$ or a limit $|1.4 - \rho - i \eta| > 0.7$.
 This still does not encroach upon the allowed region in Fig.~2, but more data
are expected. 

The process $K_L \to \pi^0 e^+ e^-$ is dominated by direct and indirect $(\sim
\epsilon)$ CP-violating contributions; there is also a small CP-conserving
two-photon contribution \cite{GB}.  The direct contribution is proportional to
$i \eta$; the indirect contribution is expected to have comparable magnitude
and the phase of $\epsilon$ (about $\pi/4$).  This process may be background
limited before the expected branching ratio ($< {\cal O}(10^{-11}$)) is
attained.  An even more challenging process, but one which would provide
valuable information on $\eta$, is $K_L \to \pi^0 \nu \bar \nu$.  The
standard-model expectation for this branching ratio is $(2.8 \pm 1.7) \times
10^{-10}$ \cite{GB}, more than 5 orders of magnitude below present upper
limits \cite{pznn}.
\bigskip

\centerline{\bf V.  CP VIOLATION IN $B$ DECAYS}
\bigskip

There is no simple analog to the $K_S$--$K_L$ system in neutral $B$ decays,
Whereas the $K_S$ and $K_L$ differ in lifetime by a factor of 600, the lifetime
differences for nonstrange $B$ mass eigenstates are expected to be negligible,
and at most 10 or 20\% for strange $B$'s \cite{Beneke}.  However, in two major
types of experiments, there should be substantial CP-violating asymmetries in
$B$ decays.  These are decays to CP eigenstates and ``self-tagging'' decays. 
\bigskip

\noindent
{\bf A.  Decays to CP eigenstates}.
Clean measurements of angles in the unitarity triangle are possible when there
is a single dominant amplitude for each process $B^0 \to f$ and $\bar B^0 \to
f$, where $f$ is a CP eigenstate.  Interference between the decay and
$B^0$--$\bar B^0$ mixing amplitudes then leads to an a difference between the
time-integrated rates $\Gamma(B^0 \to f)$ and $\Gamma(\bar B^0 \to f)$.
When $f = J/\psi K_S$, the rate asymmetry is proportional to $\sin(2 \beta)$
{\it as long as there is no additional source of $B$--$\bar B$ mixing}.
When $f = \pi^+ \pi^-$, the rate asymmetry is approximately proportional to
$\sin(2 \alpha)$ in the limit that penguin contributions can be neglected; in
order to sort these out many techniques have been suggested, including an
isospin analysis based on also detecting the $\pi^\pm \pi^0$ and $\pi^0 \pi^0$
final states \cite{pipi}.

In decays to CP eigenstates the flavor of the decaying state at time of
production must be identified by independent means, since the final state could
have come from either $B$ or $\bar B$.  We shall comment briefly on one aspect
of progress in ``tagging'' the produced $B$ in Section VIII. In
electron-positron collisions at the $\Upsilon(4S)$ the time-dependent asymmetry
is odd in the difference between the decay times of the $B$ and $\bar B$,
and would vanish upon integration over all times.  One must thus resolve these
two times, requiring either precise vertex discrimination or an asymmetric
energy configuration.  The former possibility is under consideration for the
CESR machine \cite{KB}, while the latter strategy is being employed in the
construction of asymmetric $B$ ``factories'' at SLAC and KEK.

A nice way to pinpoint the angle $\beta$ from the rate asymmetries in $B \to
J/\psi + X$ was mentioned at this Workshop \cite{KS}.  If $X = K_S$, the rate
asymmetry involves an interference between amplitudes proportional to $\cos
\beta$ and ones proportional to $\sin \beta$, leading to an asymmetry $\sim
\cos \beta \sin \beta \sim \sin(2 \beta)$.  There is considerable ambiguity in
learning $\beta$ from $\sin(2 \beta)$. However, if $X = \pi \ell \nu$ the
interference term between $K_S$ and $K_L$ contributions to this final state
(with lifetime dependence $e^{-(\gamma_S + \gamma_L)\tau/2}$) contains a
contribution $~\sim \cos(2 \beta)$, which is helpful in resolving this
ambiguity.  Since $\pi \ell \nu$ events are spread out over a kaon proper
lifetime $\tau_K \le {\cal O}(\tau_L) \simeq 600 \tau_S$, the statistical power
of the interference term is diluted with respect to the $\sin(2 \beta)$ term by
a factor of $\sim 1/300$, requiring high luminosity (which may well be
available in second-generation experiments). 
\bigskip

\noindent
{\bf B.  ``Self-tagging'' decays}.
Consider the decays $B \to f$ and $\bar B \to \bar f$, where $f$ and $\bar f$
are charge-conjugates of one another which can be distinguished experimentally.
Suppose there are two decay amplitudes contributing to $B \to f$ and $\bar B
\to \bar f$:
\beq
A(B \to f) = a_1 e^{i(\phi_1 + \delta_1)} + a_2 e^{i(\phi_2 + \delta_2)}~~,~~~
A(\bar B \to \bar f) = a_1 e^{i(- \phi_1 + \delta_1)} + a_2 e^{i(-\phi_2 +
\delta_2)}~~~,
\eeq
where $\phi_i$ are weak phases and $\delta_i$ are strong phases.  Note that
under charge conjugation, the weak phases change sign, whereas the strong
phases do not.  The rate asymmetry $A \equiv [\Gamma(f) - \Gamma(\bar f)]/
[\Gamma(f) + \Gamma(\bar f)]$ is then proportional to $\sin(\phi_1-\phi_2)
\sin(\delta_1-\delta_2)$, requiring nonzero weak {\it and} strong phases
differences in order to be observed.  Once this criterion is satisfied, it
suffices to compare total rates (or branching ratios) for the processes
$B \to f$ and $\bar B \to \bar f$.  As we shall see in Section VI, one can
sometimes learn about weak phases even in the {\it absence} of any observed
CP-violating rate asymmetry.
\bigskip

\centerline{\bf VI.  ASPECTS OF RATE MEASUREMENTS}
\bigskip

\noindent
{\bf A.  Pocket guide to direct CP asymmetries}.
In the discussion of Sec.~V B, let us suppose that both strong and weak phase
differences are non-negligible, so that $\sin(\phi_1-\phi_2) \sim \sin(\delta_1
-\delta_2) \sim {\cal O}(1)$.  Then the rate asymmetry is
\beq
A = {\cal O} \left( \frac{a_1 a_2}{a_1^2 + a_2^2} \right) \sim \frac{a_2}{a_1}
\sim \sqrt{ \frac{N_2}{N_1} }
\eeq
for $a_2 \ll a_1$.  Here $N_i = {\rm const}|a_i|^2$ is the rate associated with
the amplitude $a_i$ acting independently.  Now, the statistical error in the
asymmetry is $\delta A \sim {\cal O}(N_1^{-1/2})$ (since one is detecting a
total of about $N_1$ events).  Hence the inverse fractional error on the
asymmetry (the significance of the measurement, in standard deviations) is
$A/\delta A \sim {\cal O}(N_2^{-1/2})$.  Thus, to see an asymmetry at a
significant level one needs the rate from the rarer amplitude (here, $a_2$) to
correspond to a significant signal. 

To search for CP asymmetries in self-tagging $B$ decays one then looks for
processes with (1) at least two contributing amplitudes, (2) a sufficiently
large rate for the {\it smaller} amplitude, (3) a weak phase difference between
the two amplitudes, and (4) a good chance for a {\it strong} phase difference
between the two amplitudes.  Although this last criterion may involve some
luck, we shall indicate some likely prospects below. 
\bigskip

\noindent
{\bf B.  Interesting levels for charmless $B$ decays}.
Current upper bounds on many branching ratios for charmless $B$ decays are a
few times $10^{-5}$, whereas the likely levels for these processes are of order
$10^{-5}$ \cite{etap}.  These set the scale of the dominant amplitudes $a_1$ in
charmless $B$ decays.  The subdominant amplitudes $a_2$ are typically of order
$ \lambda a_1$, so the corresponding branching ratios are of order $\lambda^2
\times 10^{-5} \simeq 5 \times 10^{-7}$.  Thus a factor of about 100 increase
in data with respect to present samples would permit the study of interference
between dominant and subdominant amplitudes in a whole host of charmless $B$
decays, assuming the most favorable strong phase difference $\delta_1 -
\delta_2$.  This same level also would permit one to learn weak phases from
decay rates even if final state phase differences vanish, as noted in one
example below. 
\bigskip

\noindent
{\bf C.  What amplitudes matter?}
The above qualitative estimate is supported by analyses based on several
amplitudes for which evidence already exists in charmless $B$ decays.

(1) The decays $B^0 \to K^+ \pi^-$ and $B^+ \to K^0 \pi^+$ have been observed
\cite{CLEOKpi} (here we do not distinguish between a process and its
charge conjugate) with branching ratios
\beq \label{eqn:neut}
\b(B^0 \to K^+ \pi^-) = (1.5^{+0.5}_{-0.4} \pm 0.1 \pm 0.1) \times 10^{-5}~~~,
\eeq
\beq \label{eqn:chgd}
\b(B^+ \to K^0 \pi^+) = (2.3^{+1.1}_{-1.0} \pm 0.3 \pm 0.2) \times 10^{-5}~~~.
\eeq
Both processes are expected to be dominated by a strangeness-changing ($\bar b
\to \bar s$) penguin amplitude $P'$, so we average their branching ratios to
obtain the estimate \cite{etap} $|P'|^2 = (1.6 \pm 0.4) \times 10^{-5}$.  Here
and subsequently we express squares of amplitudes in units of $B$ branching
ratios. Then since $P' \sim V^*_{tb} V_{ts}$ while the strangeness-preserving
($b \to d$) penguin amplitude $P$ involves the corresponding factor $V^*_{tb}
V_{td}$, we have $|P/P'| \simeq |V_{td}/V_{ts}| \simeq \lambda$.  Thus one
expects $|P|^2 \simeq \lambda^2 |P'|^2 \simeq |P'|^2/20 \simeq 8 \times
10^{-7}$. 

(2) The decays $B^+ \to K^+ \eta'$ and $B^0 \to K^0 \eta'$ have been observed
\cite{CLEOeta} with respective branching ratios of $(6.5^{+1.5}_{-1.4} \pm
0.9) \times 10^{-5}$ and $(4.7^{+2.7} _{-2.0} \pm 0.9) \times 10^{-5}$.  Since
these are likely to be dominated by an isospin-singlet transition, we average
the two values to obtain $\b(B \to K \eta') = (5.9 \pm 1.4) \times 10^{-5}$. 
These processes are likely to receive an important contribution \cite{GReta}
from a $\bar b \to \bar s$ penguin process which contributes exclusively to
flavor-singlet meson production, as illustrated in Fig.~3.  Without the
additional contribution of this graph, one would predict \cite{etap}
$\b(B \to K \eta') = (2.4 \pm 0.6) \times 10^{-5}$.  Depending on the relative
phase of the amplitudes $S'$ and $P'$, one will have $|S'|^2$ in the range of
one to several times $10^{-5}$.  By reasoning similar to that for the
estimate of $|P/P'|$, we then find $|S|^2 = {\cal O}(5 \times 10^{-7})$.

\begin{figure}
\centerline{\epsfysize = 1.6in \epsffile {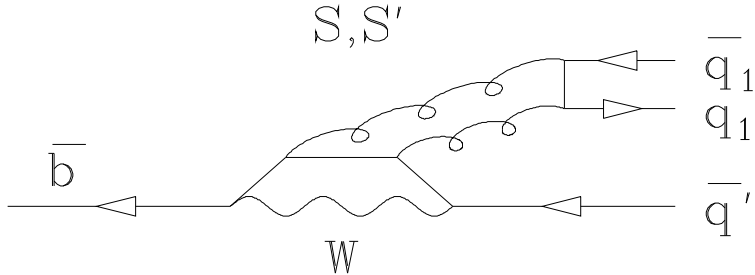}}
\caption{Example of graph contributing to amplitudes $S~(\bar q' = \bar d)$ and
$S'~(\bar q' = \bar s)$, where $q_1 \bar q_1$ stands for a flavor-SU(3)-singlet
meson.}
\end{figure}

(3) Although there is no significant evidence for either $B^0 \to \pi^+ \pi^-$
or $B^+ \to \pi^+ \pi^0$ so far, both are expected to be dominated by a
``tree'' ($T$) subprocess of the form $\bar b \to \bar u W^{*+} \to \bar u u
\bar d$, and averaging the expected contributions for the two processes using
what meager data exist \cite{CLEOKpi} we find \cite{etap} $|T|^2 \simeq 0.8
\times 10^{-5}$, an estimate which is consistent with the observed rate for $B
\to \pi \ell \nu_\ell$ \cite{LGsl}.  Then the corresponding
strangeness-changing ``tree'' amplitude $T'$, describing the subprocess $\bar b
\to \bar u W^{*+} \to \bar u u \bar s$, has square $|T'|^2 \simeq \lambda^2
|T|^2 \simeq 4 \times 10^{-7}$.  (Flavor-SU(3) breaking is likely to multiply
this estimate by a factor of $(f_K/f_\pi)^2 \simeq 1.5$.) 

Thus, there are several pieces of evidence that when branching ratios of a
few times $10^{-7}$ correspond to observable rates, one will be able to study
the interference between dominant and subdominant amplitudes in a whole range
of charmless $B$ decays.
\bigskip

\noindent
{\bf D.  Phases of amplitudes}.
\begin{table}
\caption{Dominant amplitudes, CKM combinations, and their expected phases in
charmless $B$ decays.}
\begin{center}
\begin{tabular}{|c|c c|c c|} \hline
& \multicolumn{2}{c|}{Tree} & \multicolumn{2}{c|}{Penguin} \\ \hline
$|\Delta S|$ & CKM combination & phase & CKM combination & phase \\ \hline
0 & $V_{ub}^* V_{ud}$ & $\gamma$ & $V_{tb}^* V_{td}$ & $- \beta$ \\
1 & $V_{ub}^* V_{us}$ & $\gamma$ & $V_{tb}^* V_{ts}$ & $\pi$ \\ \hline
\end{tabular}
\end{center}
\end{table}

The amplitudes expected to dominate charmless $B$ decays are summarized briefly
in Table 1. The ``tree'' amplitude is expected to dominate the $\Delta S = 0$
charmless $B$ decays such as $B^0 \to \pi^+ \pi^-$ through the combination
$V_{ub}^* V_{ud}$, while the ``penguin'' amplitude is expected to dominate the
$|\Delta S| = 1$ charmless $B$ decays such as $B^0 \to K^+ \pi^-$ through the
combination $V_{tb}^* V_{ts}$ or possibly $V_{cb}^* V_{cs}$.  The relative
phase of the tree and penguin amplitudes in the $\Delta S = 0$ decays is
$\gamma + \beta = \pi - \alpha$, while that in the $|\Delta S| = 1$ processes
is just $\gamma$.  By studying a combination of $B^0 \to \pi^+ \pi^-$, $B^0 \to
K^+ \pi^-$, $B^+ \to K^0 \pi^+$, and their charge-conjugates, one can learn
$\alpha$, $\gamma$, and the relative strong phases of tree and penguin
amplitudes \cite{DGR}. 
\bigskip

\noindent
{\bf E.  Flavor SU(3):  An application to decays with $\eta$, $\eta'$}.
Flavor SU(3) may be used in conjunction with the amplitudes discussed in
Sec.~VI C to identify processes for which there is a good chance of seeing
large effects of direct CP violation.  One is looking for processes which
involve two amplitudes of comparable strength.  The decays $B^\pm \to \pi^\pm
\eta$ and $B^\pm \to \pi^\pm \eta'$ turn out to be likely candidates
\cite{etap}. 

We performed a flavor-SU(3) decomposition \cite{DGR,ZepSWChau,GHLR,DGReta} of
decays of $B$ mesons to two light pseudoscalar mesons \cite{etap}, including
contributions from electroweak penguin terms \cite{EWP}. We neglected all
annihilation- and exchange-type amplitudes, which are expected to be highly
suppressed in comparison with those considered. We calculated expected squares
of contributions of individual amplitudes to decays, ignoring for present
purposes any interference between tree ($t,t')$ and other amplitudes. We
considered two possibilities for the relative phase of the two predominant
amplitudes, $p'$ and $s'$, in the decay $B^+ \to K^+ \eta'$, corresponding to
constructive interference and no interference between these amplitudes.  (The
amplitudes denoted by small letters are are related to those with large letters
in Sec. VI C by the inclusion of small electroweak penguin contributions.) 

We found that the branching ratios for $B^+ \to \pi^+ \eta$ due to the
($|t|^2$, $|p|^2$) contributions acting alone are (2.8,1) $\times 10^{-6}$,
within a factor of 3 of one another.  Aside from small electroweak penguin
corrections, the weak phases of $t$ and $p$ amplitudes are expected to differ
by $\alpha$ (mod $\pi$) as noted in Table 1.  Thus, if the strong phases also
differ appreciably, there is a good chance for a sizeable rate difference
between $B^+ \to \pi^+ \eta$ and $B^- \to \pi^- \eta$. 

A similarly optimistic conclusion may be drawn for $B^\pm \to \pi^\pm \eta'$.
The $|t|^2$ contribution to the branching ratio was found to be $1.4 \times
10^{-6}$, with the $|s|^2$ contribution of the same order.  The relative weak
phase of the two amplitudes is again $\alpha$ (mod $\pi$), while if the $|s|^2$
contribution is due to rescattering or other long-distance effects \cite{BFC}
the relative strong phase could be appreciable.
\bigskip

\noindent
{\bf F.  $B \to$ Vector ($V$) + Pseudoscalar ($P$) decays}.  
The CLEO Collaboration \cite{CLEOVP} has now observed the decays $B^+ \to
\omega \pi^+$ and $B \to \omega K^+$ with branching ratios of
$(1.1^{+0.6}_{-0.5} \pm 0.2) \times 10^{-5}$ (2.9$\sigma$) and
$(1.5^{+0.7}_{-0.6} \pm 0.3) \times 10^{-5}$ (4.3$\sigma$), respectively. These
are the first reported decays of $B$ mesons to charmless final states involving
a vector ($V$) and a pseudoscalar ($P$) meson.  $VP$ final states may be
crucial in studies of CP violation in $B$ decays \cite{NQ}. 

In Ref.~\cite{VP} we used flavor SU(3) and the observed decays to anticipate
the observability of other charmless $B \to VP$ decays in the near future.  We
identified amplitudes in the flavor-SU(3) decomposition likely to be large as a
result of present evidence. These consist of a strangeness-preserving ``tree''
amplitude $t_V$ and a strangeness-changing penguin amplitude $p'_V$.  In both
cases the subscript indicates that the spectator quark is incorporated into a
vector ($V$) meson. Other decays depending on the amplitude $t_V$ are $B^+ \to
\rho^0 \pi^+$ and $B^0 \to \rho^- \pi^+$.  If $t_V$ is the dominant amplitude
in these processes, we expect $\Gamma(B^+ \to \rho^0 \pi^+) = \Gamma(B^+ \to
\omega \pi^+)$ and $\Gamma(B^0 \to \rho^- \pi^+) = 2 \Gamma(B^+ \to \omega
\pi^+)$.  Furthermore, model calculations predicting $|t_P| > |t_V|$ imply that
decays expected to be dominated by $t_P$, such as $B^+ \to \rho^+ \pi^0$ and
$B^0 \to \rho^+ \pi^-$, will also have branching ratios in excess of $10^{-5}$.

An appreciable value for the amplitude $p'_V$, somewhat of a surprise in
conventional models \cite{VPmodels}, implies that $B \to \rho K$ decays should
be observable at branching ratio levels in excess of $10^{-5}$.  The smallness
of the ratio ${\cal B}(B^+ \to \phi K^+)/{\cal B}(B^+ \to \omega K^+)$
indicates that $|p'_P| < |p'_V|$.  The amplitude $p'_P$ should dominate not
only $B \to \phi K$ but also $B \to K^* \pi$ decays.  Evidence for any of these
would then tell us the magnitude of $p'_P$.  The relative phase of $p'_P$ and
$p'_V$ is probed by $B \to K^* (\eta,\eta')$ decays. 

Once the dominant amplitudes have been determined, flavor SU(3) predicts the
remaining tree and penguin amplitudes. One can then (cf. Ref.~\cite{etap})
determine which processes are likely to exhibit noticeable interferences
between two or more amplitudes, thereby having the potential for displaying
direct CP-violating asymmetries.  These asymmetries should be visible once one
has attained about a factor of 100 increase in data over present samples:  a
factor of 5 to reach the expected ${\cal O}(10^{-5})$ branching ratios for
dominant amplitudes, and a further factor of $\lambda^{-2} \simeq 20$ to see
the subdominant rates (as in Secs.~VI A and B). 
\bigskip

\noindent
{\bf G.  Statistical requirements for determining $\gamma$ in $B^\pm \to D
K^\pm$ decays}.
Gronau and Wyler \cite{GW} have pointed out that one can measure the angle
$\gamma$ if one measures the rates for $B^+ \to D^0 K^+$, $B^+ \to \bar D^0
K^+$, $B^+ \to D_{CP}^0 K^+$ (where $D_{CP}^0$ is a CP eigenstate), and the
corresponding charge-conjugate processes.  (Atwood, Dunietz, and Soni
\cite{ADS} have noted that this method requires one to take account of
interference between CKM-favored and doubly suppressed decays of the neutral
charmed meson.)  The rarest of these decays is the color-suppressed $B^+ \to
D^0 K^+$, whose branching ratio is probably a few $\times 10^{-6}$, or
\cite{StoneCS} $(\lambda/3)^2$ times that of the color-suppressed process $B^+
\to J/\psi K^+$, whose branching ratio is about $10^{-3}$ \cite{PDG}.  With an
effective 10\% detection efficiency for $D^0$ one is again at an effective
branching ratio of a few times $10^{-7}$, as in the charmless $B$ decays
mentioned earlier.
\bigskip

\noindent
{\bf H.  Measuring $\gamma$ in $B \to K \pi$}.
The decay $B^+ \to K^0 \pi^+$ is expected to be dominated by the $\bar b \to
\bar s$ penguin amplitude in the limit that rescattering (or annihilation)
contributions \cite{rescatt} can be neglected \cite{GHLR}.  In this case, which
we assume, the rates for $B^+ \to K^0 \pi^+$ and $B^- \to K^0 \pi^-$ should be
equal. Other tests for the presence of rescattering effects have been proposed
\cite{BGR}. 

The decays $B^0 \to K^+ \pi^-$ and $\bar B^0 \to K^- \pi^+$, on the other hand,
are dominated by the penguin amplitude but should receive some contribution
from the tree subprocess as well.  One then forms the ratio
\beq 
R \equiv \frac{\Gamma(K^+ \pi^-) + \Gamma(K^- \pi^+)}{\Gamma(K^0 \pi^+)
+ \Gamma(\bar K^0 \pi^-)} = 1 - 2 r \cos \gamma \cos \delta + r^2~~~,
\eeq
where $r \equiv |T'/P'|$ and $\delta$ is a strong phase difference between
the penguin and tree amplitudes \cite{RFa,FM,RFb}.  Fleischer and Mannel
\cite{FM} have pointed out that if $R <1$, a useful bound $\sin^2 \gamma <
R$ holds for any $r$ and $\delta$.  At present $R = 0.65 \pm 0.40$
\cite{CLEOKpi}, so the central value of $R$ is indeed below 1, but no conclusion
can be reached yet. 

If the ratio $r$ is known, one can do better \cite{GRKpi}.  One can combine
the above ratio with the CP-violating rate ``pseudo-asymmetry''
\beq
A_0 \equiv \frac{\Gamma(B^0 \to K^+ \pi^-) - \Gamma(\bar B^0 \to 
K^- \pi^+)} {\Gamma(B^+ \to K^0 \pi^+) + \Gamma(B^- \to \bar K^0 \pi^-)}~~~, 
\eeq
to find an expression for $\gamma$ which depends only on these quantities and
on the ratio $r$, for which we provide an estimate based on $B \to \pi \pi$ and
$B \to \pi \ell \nu_\ell$ decays. A similar idea can be applied to $B_s \to K^+
K^-$ and $B_s \to K^0 \bar K^0$ decays \cite{RFb,DF}. 

\begin{figure}
\centerline{\epsfysize = 4 in \epsffile {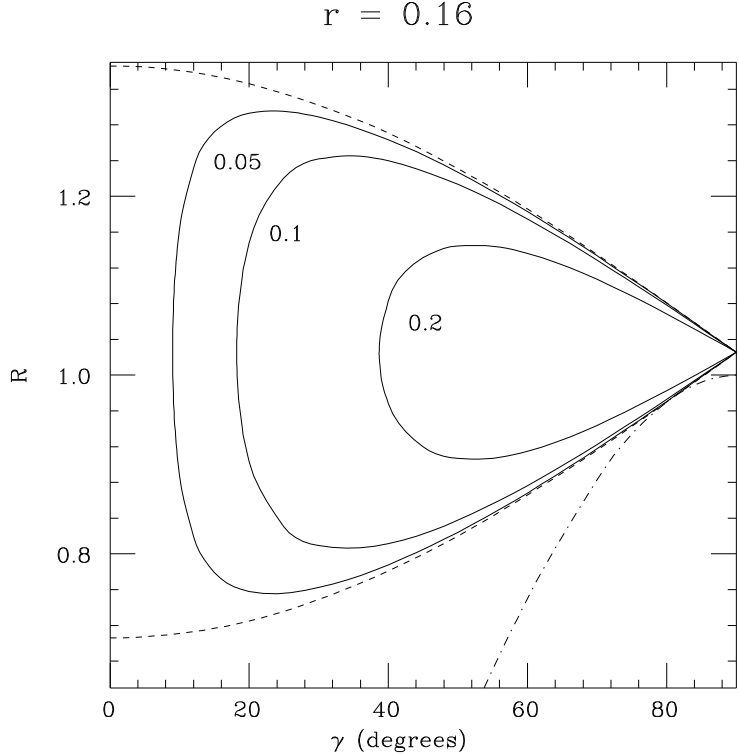}}
\caption{Value of $R$ (ratio of neutral to charged $B \to K \pi$ partial
widths) as a function of $\gamma = {\rm Arg}(V^*_{ub})$ for $r=0.16$.  Solid
lines are labeled by values of pseudo-asymmetry parameter $|A_0|$.  Dotted
boundary lines correspond to $A_0=0$.  Also shown (dot-dashed lines) is
the Fleischer-Mannel bound $\sin^2 \gamma \leq R$.} 
\end{figure}

In Fig.~4 we plot contours of fixed asymmetry $|A_0|$ in the $\gamma$--$R$
plane. (Note that one cannot distinguish between $\gamma$ and $\pi - \gamma$
using this method since when $\gamma \to \pi -\gamma,~\delta \to \pi - \delta$
both $R$ and $A_0$ are unchanged.) The case $r = 0.16$ is shown; for other
values see Ref.~\cite{GRKpi}.  Present data imply errors on $r$ of $\pm 0.06$,
about a factor of 4 too large to permit a useful determination of $\gamma$.
Assuming $r$ is sufficiently well known, that $45^\circ < \gamma < 135^\circ$,
that rescattering effects and electroweak penguins are negligible (they may not
be; see Ref.~\cite{GRKpi} for details), a measurement of $\gamma$ to $\pm
10^\circ$ requires $R$ to be known to $\pm 0.028$, or a data sample of about
200 times the present one \cite{CLEOKpi} of 3.3 million $B \bar B$ pairs.  The
determination of $\gamma$ to $\pm 10^\circ$ and $\eta$ to $\pm 0.02$ (mentioned
in Sec.~III D) can significantly restrict the range of parameters in Fig.~2. 
\bigskip

\centerline{\bf VII.  CHARM-ANTICHARM ANNIHILATION}
\bigskip

A number of features of $B$ decays look just enough out of line with respect
to theoretical expectations to be worth a raised eyebrow.

\begin{enumerate}

\item The semileptonic branching ratio $\b(B \to X \ell \nu)$ is about 11\%
(vs.~a theoretical prediction of about 12\%) \cite{MN}.

\item The number $n_c$ of charmed particles per average $B$ decay is about
1.1 to 1.2 vs.~a theoretical prediction of 1.2 to 1.3 \cite{MN}.

\item The inclusive branching ratio $\b(B \to \eta' X)$ appears large
\cite{CLEOeta} in comparison with theoretical expectations \cite{incl}.

\item The exclusive branching ratio $\b(B \to K \eta')$ appears to require an
additional contribution (as illustrated in the example of Fig.~3) in
comparison with the penguin contribution leading to $B^0 \to K^+ \pi^-$ or $B^+
\to K^0 \pi^+$. 

\end{enumerate}

A common source for these effects could be an enhanced rate for the subprocess
$\bar b \to \bar c c s \to \bar q q \bar s$, where $q$ stands for a light quark
\cite{Dunietz}, e.g., through rescattering effects.  These are inherently
long-range and nonperturbative and could also be responsible for the overall
enhancement of the $\bar b \to \bar s$ penguin transitions noted in
Refs.~\cite{BFC}. Alternatives for points (3) and (4) which have been suggested
include a large $c \bar c$ \cite{KBe} or gluonic component in the $\eta'$. The
former possibility is intriguing but one must then ascribe the suppression of
the decay $J/\psi \to \eta' \gamma$ to form factor effects.  A large gluonic
admixture in the $\eta'$ would depress the predicted branching ratio $\b(\phi
\to \eta' \gamma)$ below the value of about $10^{-4}$ predicted \cite{JLRetap}
if $\eta'$ is mainly a $q \bar q$ state.  At this Workshop I have learned that
the CMD-2 Detector at VEPP-2 in Novosibirsk has presented the first evidence (6
events) for the long-sought $\phi \to \eta' \gamma$ decay \cite{CMD}, with a
branching ratio consistent with the standard prediction. 

If rescattering from the $\bar b \to \bar c c \bar s$ subprocess into states
containing light quarks really is important, both the penguin amplitude $P'$
and the singlet penguin $S'$ mentioned in Sec. VI C could have strong phases
very different from the tree amplitude $T'$, raising the possibility of
substantial CP-violating asymmetries whenever these amplitudes interfere with
one another in a self-tagging $B$ decay (such as $B^0 \to K^+ \pi^-$). 
\bigskip

\centerline{\bf VIII.  TAGGING REMARKS}
\bigskip

\noindent
{\bf A.  Same-side tagging}.  Several years ago a method was proposed
for identifying the flavor of a neutral $B$ at the time of production by means
of correlations with the charge of the pion produced nearby in phase space
\cite{tags}.  There has been considerable progress in utilizing this method,
originally at LEP \cite{LEPtags} but more recently by the CDF Collaboration
\cite{Schmidt}.  The idea is simple, and is even incorporated into existing
Monte Carlo fragmentation programs:  A produced $\bar b$, when fragmenting
into a $B^0$, requires production of a $d$ quark, so that a $\bar d$ quark is
the next quark down the rapidity chain.  If this quark is incorporated into a
charged pion, that pion must be a $\pi^+$.  Thus, one expects a correlation
between $\pi^+$ and $B^0$, and, correspondingly, between $\pi^-$ and $\bar
B^0$. 

If the fragmentation of a leading $b$ or $\bar b$ quark takes place in a
charge-independent manner, one should expect the $B^+ \pi^-$ correlation to
be the same as the $B^0 \pi^+$ correlation, by isospin reflection.  This is
not what CDF sees; the $B^+ \pi^-$ correlation tends to be stronger.
Several reasons for this have been noted \cite{DRtag}.  One instrumental effect
against which one must guard is the misidentification of charged kaons as
charged pions.  A correlation is expected between charged $B$'s and oppositely
charged kaons, but not between neutral $B$'s and charged kaons.
\bigskip

\noindent
{\bf B.  Importance of resonances}.  The $B^0 \pi^+$ and $\bar B^0 \pi^-$
correlations mentioned above are just those expected from resonance decays.
Now, decays of narrow resonances can lead to a considerable improvement of
the signal-to-noise ratio in studying such correlations. In the case of charm,
the vector meson decays $D^{*+} \to D^0 \pi^+$ and $D^{*-} \to \bar D^0 \pi^-$
have been crucial in identifying neutral $D$'s, since the pions are almost at
rest with respect to them.  The corresponding $B^*$ vector mesons lie too close
in mass to the $B$'s for a strong decay. One must look to the next excited
states, consisting of a $b$ and a light antiquark ($\bar q$) in a P-wave. 

There are two known P-wave $c \bar q$ resonances \cite{PDG}:  the $D^*_2(2460)$
and the $D_1(2420)$, where the subscript stands for the total angular momentum
$J$.  These decay via D-wave final states:  $D^*_2 \to [(D~{\rm or}~D^*) \pi]
_{\ell=2}$, $D_1 \to [D^* \pi]_{\ell=2}$ and hence are fairly narrow. 
Candidates for the corresponding $B^*_2(\sim 5740)$ and $B_1(\sim 5730)$ states
have been seen \cite{LEPtags}.  Such states should be useful in tagging neutral
$B$'s, as has been emphasized in Refs.~\cite{EHQ}.

One is still missing two predicted P-wave resonances in both the $c \bar q$ and
$b \bar q$ systems.  One, with $J=1$, should decay to $[(D^*~{\rm or}~B^*)
\pi]_{\ell = 0}$, while the other, with $J=0$, should decay to $[(D~{\rm or}~B)
\pi]_{\ell = 0}$.  These could in part be responsible for the observed
correlations, since they are expected to be broad (being able to decay via
S-waves) and hence difficult to distinguish from nonresonant effects.
\bigskip

\centerline{\bf IX.  EFFECTS BEYOND THE STANDARD MODEL}
\bigskip

\noindent
{\bf A. Loop-dominated $B$ decays.}
Table 2 summarizes some $B$ decay processes which are dominated by loop
diagrams.  The theoretical numbers are taken from the indicated references
or from a recent review by Ali \cite{Aliloops}.  The experimental results are
from Refs.~\cite{CDFmumu}--\cite{Skw}. New particles in loops can include
supersymmetric partners, extra Higgs bosons, new quarks, new $q \bar q' t \bar
t$ interactions, and much more \cite{HW,Burdman}. 

\begin{table}
\caption{Decays of nonstrange $B$ mesons dominated by loop diagrams.}
\begin{center}
\begin{tabular}{c c c} \hline
Decay & Experimental limit & Branching ratio in \\
Mode  &      or rate       &   standard model \\ \hline
$ \mu^+ \mu^-$ & $< 6.9 \times 10^{-7~a}$ & $1.1 \times 10^{-10~b}$ \\
$ \gamma \gamma$ & $< 3.9 \times 10^{-5~c}$ & $10^{-8}$ \\
$ X_s \gamma$ & $(2.32 \pm 0.57 \pm 0.35) \times 10^{-4~d}$ & \\
              & $(3.29 \pm 0.71 \pm 0.68) \times 10^{-4~e}$ & 
 $(3.25 \pm 0.30^{f} \pm 0.40) \times 10^{-4~g}$ \\
$ X_s \mu^+ \mu^-$ & $< 4.7 \times 10^{-5~h}$ & $(5.73^{+0.75}_{-0.78})
\times 10^{-6}$ \\
$K^0 (e^+e^-/\mu^+ \mu^-)$ & $< (1.5/2.6) \times 10^{-4~i}$ &
 $([5 \pm 3]/[3 \pm 1.8]) \times 10^{-7}$ \\
$K^+ (e^+e^-/\mu^+ \mu^-)$ & $< (1.2/0.9) \times 10^{-4~i}$ &
 $([5 \pm 3]/[3 \pm 1.8]) \times 10^{-7}$ \\ \hline

\end{tabular}
\end{center}
\leftline{$^a$~CDF \protect\cite{CDFmumu}.  $^b$~Rate $\sim m_\ell^2$ for small
lepton mass $m_\ell$. $^c$~L3 \protect\cite{L3gg}. $^d$~CLEO
\protect\cite{CLEOsg}.}
\leftline{$^e$~ALEPH \protect\cite{ALsg} $^f$~Scale, $m_t$ error. 
$^g$~Ref.~\protect\cite{sgth}. $^h$~CLEO \protect\cite{Skw}, $\ell^+ \ell^-$.
$^i$~CLEO \protect\cite{Skw}.} 
\end{table}

In the standard model, the loop diagrams affecting $b \to s \gamma$ involve an
intermediate $u,c,t$ and an intermediate $W^-$.  These also contribute to $b
\to s \ell^+ \ell^-$, with the $\ell^+ \ell^-$ pair produced by a virtual
photon or $Z$.  However, the process $b \to s \ell^+ \ell^-$ also receives a
contribution from a box diagram with an intermediate $W^+ W^-$ pair.  Thus,
non-standard physics can affect $b \to s \gamma$ and $b \to s \ell^+ \ell^-$
in different ways.  The ratio of rates for these two processes, the
$m_{\ell^+ \ell^-}$ spectrum, and the energy distributions of the leptons are
all tools which can be used to distinguish among non-standard models.
\bigskip

\noindent
{\bf B.  Supersymmetry}.
Loop diagrams can contain not only superpartners, but also additional Higgs
bosons.  A 25\% comparison between theory and experiment can lead to
significant restrictions on the parameter space \cite{Grant}.
\bigskip

\noindent
{\bf C. Technicolor-like interactions}.
There may be new higher-dimension $\bar b s \bar t t$ interactions which, when
integrated over the $t \bar t$ loop, contribute to $b \to s \gamma$.  There can
also be modifications of the rates for $b \to s \ell^+ \ell^-$ and $b \to s \nu
\bar \nu$.  The latter may be correlated with modifications to $K \to \pi \nu
\bar \nu$ rates \cite{Burdman}. 
\bigskip

\noindent
{\bf D. New quark families}.
The invisible decays of the $Z$ indicate that there are only three families
of light neutrinos with standard electroweak couplings.  However, the
possibility still remains open of additional families of quarks and leptons
if the corresponding neutrinos are sufficiently heavy, or of exotic quarks
beyond the standard left-handed doublets $(u,d),(c,s),(t,b)$.  Thus, one can
have modifications of $V_{ts}$ and $V_{td}$, more $Q=2/3$ quarks in loops,
or even anomalous $t \bar t \gamma$ and $t \bar t Z$ couplings.  All these
can affect $b \to s \gamma$ and $b \to s \ell^+ \ell^-$ processes.
\bigskip

\noindent
{\bf E. New Higgs bosons}.
Supersymmetry and non-supersymmetric grand unified theories beyond SU(5) both
predict an extended Higgs boson sector, with charged Higgs bosons that can
participate in loop diagrams.  Since these bosons in general couple differently
than $W^\pm$, they can affect the relative rates of $b \to s \gamma$ and $b \to
s \ell^+ \ell^-$, and the kinematic variables in the latter.
\bigskip

\noindent
{\bf F. $B \to \phi K_S$ and new physics}.
A very pretty observation \cite{GWo} illustrates a way in which new physics can
show up in decay amplitudes.  One compares the decay $B^0 \to \phi K_S$, which
is expected to be dominated by the $\bar b \to \bar s$ penguin amplitude (with
weak phase $\pi$) with the decay $B^0 \to J/\psi K_S$, which is expected to be
dominated by the subprocess $\bar b \to \bar c c \bar s$, with weak phase 0.
The CP-violating asymmetry in both these decays arise from the interference
of the decay amplitude with the $B^0$--$\bar B^0$ mixing amplitude (whose weak
phase is $2 \beta$).  Thus the standard model predicts a time-integrated
rate asymmetry
\beq
\frac{\Gamma(B^0 \to \phi K_S) - \Gamma(\bar B^0 \to \phi K_S)}
{\Gamma(B^0 \to \phi K_S) + \Gamma(\bar B^0 \to \phi K_S)} =
\frac{\Gamma(B^0 \to J/\psi K_S) - \Gamma(\bar B^0 \to J/\psi K_S)}
{\Gamma(B^0 \to J/\psi K_S) + \Gamma(\bar B^0 \to J/\psi K_S)} =
- \frac{x \sin(2 \beta)}{1+x^2}
\eeq
where $x \equiv [\Delta m / \Gamma]_{B^0} \simeq 0.7$.  New physics in the
penguin amplitude can change the weak phase in $B^0 \to \phi K^0$, so that the
asymmetries in $\phi K_S$ and $J/\psi K_S$ are no longer equal.  Standard
sources in fact can lead to a difference between the two asymmetries of at
most a few \%; differences in excess of this figure are considered interesting
\cite{GWo}.  The absence of an anomalously high $B^+ \to \phi \pi^+$ rate
(the current branching ratio upper limit \cite{CLEOVP} is $\b < 5 \times
10^{-6}$) and an upper bound on $\b(B^+ \to \bar K^{*0} K^+)$ are useful in
bounding any anomalously high contribution of the $u \bar u$ loop in the
$\bar b \to \bar s$ penguin diagram.
\bigskip

\centerline{\bf X.  SUMMARY}
\bigskip

\noindent
{\bf A.  What lies ahead?}
(1) Are the CKM matrix and its phases the source of CP violation for $K_L$
and $K_S$?  If so, we may see a deviation in the double ratio (\ref{eqn:Krat})
from unity.  (2) Do $B$ decays provide a consistent set of CKM phases?  If so,
there are definite predictions for the time-integrated rate asymmetry in the
$J/\psi K_S$ final state.  Progress has been made in ``tagging'' neutral
$B$'s, necessary to identify this asymmetry.  One expects many checks of the
CKM predictions in measurements of decay rates of charged and neutral $B$'s.

A ratio of exactly 1 for (\ref{eqn:Krat}) unfortunately lies within the allowed
parameter space for the CKM theory.  However, the potential for discrepancies
with respect to the CKM theory is very great in measurements using $B$ mesons. 
If such discrepancies are found, one will have to explore other sources of CP
violation, such as superweak models, right-handed $W$'s, multi-Higgs models,
and supersymmetry.  A generic feature of these last three is the prediction of
electron and neutron electric dipole moments \cite{Barr} not far below present
limits; the CKM theory predicts values well below these and probably
inaccessible to foreseeable experiments. 
\bigskip

\noindent
{\bf B. Quark and lepton families.}
Another possibility for experiments with $B$'s is that they will fail to
expose any inconsistencies in the CKM picture.  This will sharpen the question
of where the CKM phases really originate; we do not understand that any better
than we understand quark and lepton masses.  Such an understanding is far
overdue and is not provided by any theories currently on the market.  The
pattern of quark and lepton masses and couplings (see, e.g., Ref.~\cite{JRKEK})
is strongly reminiscent of a level structure, with intensity rules favoring
nearest-neighbor transitions as in atomic physics or quarkonium \cite{dipole}.
\bigskip

\noindent
{\bf C.  Favorite nonstandard model.}
For one example of physics beyond the standard model one can look to grand
unified groups beyond SU(5) and SO(10), such as $E_6$ (see, e.g,
Ref.~\cite{ES}).  Each family of quarks and leptons belongs to a reducible
${\bf 5^*} + {\bf 10}$ multiplet of SU(5), which becomes a ${\bf 16}$-spinor of
SO(10) [$= {\bf 5^*} + {\bf 10} + {\bf 1}$ of SU(5)] when a right-handed
neutrino is added.  The Higgs boson naturally belongs to a vector ${\bf 10}$
representation of SO(10), which in supersymmetry would be accompanied by a
${\bf 10}$ of fermions.  The simplest representation containing both a
${\bf 16}$ and a ${\bf 10}$ of fermions, once we add a further ${\bf 1}$ of
fermions, is the ${\bf 27}$-plet of $E_6$.  Now, if we like, we can throw away
the supersymmetry and simply consider the properties of matter in the ${\bf
27}$-plet.  The ${\bf 10}$-plet of SO(10) decomposes into a ${\bf 5} + {\bf
5^*}$ of SU(5), consisting of the following left-handed particles and their
charge-conjugates:  (1) an isoscalar, color-triplet quark $h$; (2) a
vector-like charged lepton $E^+$, and (3) the corresponding (anti)neutrino
$\nu_E$.  The ${\bf 1}$ of SO(10) is a sterile left-handed neutrino.  Each
${\bf 16}$-plet family of quarks and leptons is accompanied by a ${\bf 10}$
and a ${\bf 1}$.  The ordinary quarks $d,s,b$ of charge $-1/3$ can mix with
the three different $h$'s, leading to a variety of flavor-changing neutral
currents.  Thus, the unitarity triangle may not close, with interesting
consequences for experiments with $B$ mesons \cite {DS}.
\bigskip

\noindent
{\bf D. Demonstrations and conclusions.}
I close with two demonstrations.  The first (see \cite{JRKEK}) is one you may
want to try on audiences who have not heard it before; show them a block of 8
by 4 squares and ask them if they recognize the pattern.  Then separate the
block into two pieces, with 2 columns of 4 squares on the left and 6 columns of
4 on the right, and ask again.  Finally fill in the missing squares --
hydrogen, helium, and the transition metals -- to form the periodic table of
the elements.  Most people will recognize it by this point; the variety of the
pattern usually is the key. A few individuals see the pattern immediately. 

The second uses an asymmetric top \cite{toy} whose principal axes of inertia
are not aligned with the axis with which it makes contact with the surface on
which it is placed.  As a result, it can spin in one direction freely, but
if it is set spinning in the other direction, it will gradually slow down and
then reverse its direction.  Question:  what symmetries are violated?  Answer:
P and T.  T violation in particle physics is likely to be more subtle!
\bigskip

\centerline{\bf XI.  ACKNOWLEDGEMENTS}
\bigskip

I would like to thank Amol Dighe, Isi Dunietz, and Michael Gronau for enjoyable
collaborations on many of the subjects mentioned here.  Some of the research
described here was performed at the Aspen Center for Physics.  This work was
supported in part by the United States Department of Energy under Grant No.~DE
FG02 90ER40560. 
\bigskip

\def \ajp#1#2#3{Am.~J.~Phys. #1 (#3) #2}
\def \ap#1#2#3{Ann.~Phys.~(N.Y.) #1 (#3) #2}
\def \apny#1#2#3{Ann.~Phys.~(N.Y.) #1 (#3) #2}
\def \app#1#2#3{Acta Physica Polonica #1 (#3) #2}
\def \arnps#1#2#3{Ann.~Rev.~Nucl.~Part.~Sci. #1 (#3) #2}
\def \arns#1#2#3{Ann.~Rev.~Nucl.~Sci. #1 (#3) #2}
\def \art{and references therein}
\def \ba88{Particles and Fields 3 (Proceedings of the 1988 Banff Summer
Institute on Particles and Fields), edited by A. N. Kamal and F. C. Khanna
(World Scientific, Singapore, 1989)}
\def \baps#1#2#3{Bull.~Am.~Phys.~Soc. #1 (#3) #2}
\def \be87{Proceedings of the Workshop on High Sensitivity Beauty
Physics at Fermilab, Fermilab, Nov. 11--14, 1987, edited by A. J. Slaughter,
N. Lockyer, and M. Schmidt (Fermilab, Batavia, IL, 1988)} 
\def \btasi{Testing the Standard Model (Proceedings of the 1990
Theoretical Advanced Study Institute in Elementary Particle Physics),
edited by M. Cveti\v{c} and P. Langacker (World Scientific, Singapore, 1991)}
\def \cmts#1#2#3{Comments on Nucl.~and Part.~Phys. #1 (#3) #2}
\def \cn{Collaboration}
\def \corn{Lepton and Photon Interactions:  XVI International Symposium,
Ithaca, NY 1993, edited by P. Drell and D. Rubin (AIP, New York, 1994)}
\def \cp89{{\it CP Violation,} edited by C. Jarlskog (World Scientific,
Singapore, 1989)} 
\def \dpfa{The Albuquerque Meeting:  DPF 94 (Division of Particles and
Fields Meeting, American Physical Society, Albuquerque, NM, August 2--6,
1994), ed. by S. Seidel (World Scientific, River Edge, NJ, 1995)}
\def \dpff{The Fermilab Meeting -- DPF 92 (Division of Particles and
Fields Meeting, American Physical Society, Fermilab, 10--14 November, 1992),
ed. by C. H. Albright \ite~(World Scientific, Singapore, 1993)} 
\def \dpfm{The Minneapolis Meeting:  DPF 96 (Division of Particles and
Fields Meeting, American Physical Society, Minneapolis, MN, 10--15 August,
1996), to be published}
\def \dpfv{The Vancouver Meeting - Particles and Fields '91
(Division of Particles and Fields Meeting, American Physical Society,
Vancouver, Canada, Aug.~18--22, 1991), ed. by D. Axen, D. Bryman, and M. Comyn
(World Scientific, Singapore, 1992)} 
\def \efi{Enrico Fermi Institute Report No.~}
\def \epj#1#2#3{Eur.~Phys.~J.~#1 (#3) #2}
\def \fermlg{Proc.~Int.~Symp.~on Lepton and Photon Interactions at High
Energies (Fermilab, August 23--29, 1979), T. B. W. Kirk and H. D. I.
Abarbanel, eds., Fermilab, Batavia, IL (1979)}
\def \hb87{Proceeding of the 1987 International Symposium on Lepton and
Photon Interactions at High Energies, Hamburg, 1987, ed. by W. Bartel
and R. R\"uckl (Nucl.~Phys.~B, Proc. Suppl., vol. 3) (North-Holland,
Amsterdam, 1988)}
\def \ib{{\it ibid.}}
\def \ibj#1#2#3{{\it ibid.} #1 (#3) #2}
\def \ijmpa#1#2#3{Int.~J. Mod.~Phys.~A #1 (#3) #2}
\def \jpb#1#2#3{J. Phys.~B #1 (#3) #2}
\def \jpg#1#2#3{J. Phys.~G #1 (#3) #2}
\def \KEK{{\it Flavor Physics} (Proceedings of the Fourth International
Conference on Flavor Physics, KEK, Tsukuba, Japan, 29--31 October 1996),
edited by Y. Kuno and M. M. Nojiri, Nucl.~Phys.~B Proc.~Suppl.~59 (1997)}
\def \kdvs#1#2#3{Kong.~Danske Vid.~Selsk., Matt-fys.~Medd. #1 (#3) No.~#2}
\def \ky{Proceedings of the International Symposium on Lepton and
Photon Interactions at High Energy, Kyoto, Aug.~19-24, 1985, edited by M.
Konuma and K. Takahashi (Kyoto Univ., Kyoto, 1985)} 
\def \latm{Lattice 1995 (Proceedings of the International Symposium on
Lattice Field Theory, Melbourne, Australia, 11--15 July 1995, T. D. Kieu,
B. H. J. McKellar, and A. Guttman, eds., North-Holland, Amsterdam, 1996}
\def \lgb{LP95: Proceedings of the International Symposium on Lepton and
Photon Interactions (IHEP), 10--15 August 1995, Beijing, People's Republic of
China, Z.-P. Zheng and H.-S. Chen, eds., World Scientific, Singapore, 1996} 
\def \lgg{International Symposium on Lepton and Photon Interactions, Geneva,
Switzerland, July, 1991}
\def \lkl87{Selected Topics in Electroweak Interactions (Proceedings of 
the Second Lake Louise Institute on New Frontiers in Particle Physics, 15--21
February, 1987), edited by J. M. Cameron \ite~(World Scientific, Singapore,
1987)}
\def \lti{lectures at this Institute}
\def \mpla #1#2#3{Mod.~Phys.~Lett. A #1 (#3) #2}
\def \nc#1#2#3{Nuovo Cim. #1 (#3) #2}
\def \nima#1#2#3{Nucl.~Inst.~Meth.~A #1 (#3) #2}
\def \np#1#2#3{Nucl.~Phys. #1 (#3) #2}
\def \npbps#1#2#3{Nucl.~Phys.~B (Proc.~Suppl.) #1 (#3) #2}
\def \npps#1#2#3{Nucl.~Phys.~(Proc.~Suppl.) #1 (#3) #2}
\def \oxf{Proceedings of the Oxford International Conference on
Elementary Particles 19/25 Sept.~1965, ed.~by T. R. Walsh (Chilton, Rutherford
High Energy Laboratory, 1966)}
\def \pascos{PASCOS 94 (Proceedings of the Fourth International
Symposium on Particles, Strings, and Cosmology, Syracuse University, 19--24
May 1994), ed.~by K. C. Wali (World Scientific, Singapore, 1995)}
\def \pbarp{AIP Conference Proceedings 357: 10th Topical Workshop on
Proton-Antiproton Collider Physics, Fermilab, May 1995, ed.~by R. Raja and J.
Yoh (AIP, New York, 1996)}
\def \pisma#1#2#3#4{Pis'ma Zh. Eksp. Teor. Fiz. #1 (#3) #2 [JETP Lett.,
#1 (#3) #4]} 
\def \pl#1#2#3{Phys.~Lett. #1 (#3) #2}
\def \pla#1#2#3{Phys.~Lett. A #1 (#3) #2}
\def \plb#1#2#3{Phys.~Lett. B #1 (#3) #2}
\def \ppmsj#1#2#3{Proc.~Phys.~Math.~Soc.~Jap. #1 (#3) #2}
\def \pnpp#1#2#3{Prog.~Nucl.~Part.~Phys. #1 (#3) #2}
\def \pr#1#2#3{Phys.~Rev. #1 (#3) #2}
\def \prd#1#2#3{Phys.~Rev. D #1 (#3) #2}
\def \prl#1#2#3{Phys.~Rev.~Lett. #1 (#3) #2}
\def \prp#1#2#3{Phys.~Rep. #1 (#3) #2}
\def \ptp#1#2#3{Prog.~Theor.~Phys. #1 (#3) #2}
\def \ptwaw{Plenary talk, XXVIII International Conference on High Energy
Physics, Warsaw, July 25--31, 1996, Proceedings edited by Z. Ajduk and A. K.
Wroblewski (World Scientific, River Edge, NJ, 1997)}
\def \rmp#1#2#3{Rev.~Mod.~Phys. #1 (#3) #2}
\def \si90{25th International Conference on High Energy Physics, Singapore,
Aug. 2-8, 1990, Proceedings edited by K. K. Phua and Y. Yamaguchi (World
Scientific, Teaneck, N. J., 1991)}
\def \slaclg{Proceedings of the 1975 International Symposium on
Lepton and Photon Interactions at High Energies, Stanford University,
Aug.~21--27, 1975, W. T. Kirk, ed., SLAC, Stanford, CA, (1975)} 
\def \slc{Proceedings of the Salt Lake City Meeting (Division of
Particles and Fields, American Physical Society, Salt Lake City, Utah, 1987),
ed. by C. DeTar and J. S. Ball (World Scientific, Singapore, 1987)}
\def \smass{Proceedings of the 1982 DPF Summer Study on Elementary
Particle Physics and Future Facilities, Snowmass, Colorado, edited by R.
Donaldson, R. Gustafson, and F. Paige (World Scientific, Singapore, 1982)}
\def \smassa{Research Directions for the Decade (Proceedings of the
1990 DPF Snowmass Workshop), edited by E. L. Berger (World Scientific,
Singapore, 1991)}
\def \smassb{Proceedings of the Workshop on $B$ Physics at Hadron
Accelerators, Snowmass, Colorado, 21 June--2 July 1994, ed.~by P. McBride
and C. S. Mishra, Fermilab report FERMILAB-CONF-93/267 (Fermilab, Batavia, IL,
1993)} 
\def \stone{B Decays, edited by S. Stone (World Scientific, Singapore,
1994)}
\def \tw{this Workshop}
\def \waw{XXVIII International Conference on High Energy
Physics, Warsaw, July 25--31, 1996, Proceedings edited by Z. Ajduk and A. K.
Wroblewski (World Scientific, River Edge, NJ, 1997)}
\def \yaf#1#2#3#4{Yad.~Fiz. #1 (#3) #2 [Sov.~J.~Nucl.~Phys.,~#1 (#3) #4]}
\def \zhetf#1#2#3#4#5#6{Zh.~Eksp.~Teor.~Fiz. #1 (#3) #2 [Sov.~Phys.~--
JETP, #4 (#6) #5]}
\def \zhetfl#1#2#3#4{Pis'ma Zh.~Eksp.~Teor.~Fiz. #1 (#3) #2 [JETP
Letters #1 (#3) #4]}
\def \zp#1#2#3{Zeit.~Phys. #1 (#3) #2}
\def \zpc#1#2#3{Zeit.~Phys.~C #1 (#3) #2}


\begin{thebibliography}{199}
\bibitem{CDFmumu} CDF \cn, G. Bauer, \tw.  The corresponding limit for strange
$B$'s is $\b(B_s^0 \to \mu^+ \mu^-) < 2.1 \times 10^{-6}$.  Both limits are
90\% c.l.
\end{thebibliography}
\end{document}